\documentclass[12pt,a4paper]{article}
\usepackage{jcappub}
\usepackage{epsfig} 
\usepackage{amsmath}
\usepackage{amsfonts}
\usepackage{graphicx}
\usepackage{amssymb}
\usepackage{ctable}

\title{Computation of the Halo Mass Function Using Physical Collapse Parameters: Application to Non-Standard Cosmologies} 

\author[1,2]{I. Achitouv}
\author[3]{C. Wagner}
\author[1,2,5]{J. Weller}
\author[4]{Y. Rasera}

\affiliation[1]{Universit\"ats-Sternwarte M\"unchen, Ludwig-Maximilians-Universit\"at M\"unchen, Scheinerstr. 1, 81679 M\"unchen, Germany}

\affiliation[2]{Excellence Cluster Universe, Boltzmannstr. 2, 85748 Garching bei M\"unchen, Germany}

\affiliation[3]{Max-Planck-Institute for Astrophysics, Karl-Schwarzschild-Str. 1, 5748 Garching bei M\"unchen, Germany}

\affiliation[4]{Max-Planck-Institute for Extraterrestrial Physics, Giessenbachstrasse, 85748 Garching, Germany}  

\affiliation[5]{Laboratoire Univers et Th\'eories (LUTh),\\ UMR 8102
CNRS, Observatoire de Paris, Universit\'e Paris Diderot, \\ 5 Place
  Jules Janssen, 92190 Meudon, France}

\emailAdd{Ixandra.Achitouv@usm.lmu.de}

\abstract{In this article we compare the halo mass function predicted
  by the excursion set theory with a drifting diffusive barrier
  against the results of N-body simulations for several cosmological
  models. This includes the standard $\Lambda$CDM case for a large
  range of halo masses, models with different types of primordial
  non-Gaussianity, and the Ratra-Peebles quintessence model of Dark Energy. We show that in all those
  cosmological scenarios, the abundance of dark matter halos can be
  described by a drifting diffusive barrier, where the two parameters
  describing the barrier have physical content. In the case of the
  Gaussian $\Lambda CDM$, the statistics are precise enough to actually
  predict those parameters at different redshifts from the initial conditions. 
Furthermore, we found that the stochasticity
in the barrier is non-negligible making the simple deterministic
spherical collapse model a bad approximation even at very high halo masses.  
We also show that using the standard excursion set approach with a
barrier inspired by peak patches leads to inconsistent predictions of
the halo mass function.} 

\begin{document}
\maketitle

\section{Introduction}

The abundance of dark matter halos is sensitive to both the initial
statistics of the matter density field
\cite{LucchinMatarrese1988,Colafrancesco1989,Chui1998,Eric00} and the
gravitational collapse, which has led to their formation
(e.g.~\cite{GunnGott1972,wangSteinhardt,Micki}). Hence, the abundance
of these virialized structures at different cosmic times is sensitive
to the underlying cosmology (e.g.~\cite{Vik,Pill,Annali}). Therefore an
accurate modelling of the halo mass function is crucial for future
observational analysis, but additionally it may provide a physical
understanding of halo formation. A successful framework to predict the
halo abundances is the excursion set theory \cite{Bondetal}. Given the
initial statistics of the matter density field and  collapse
criteria, one can compute the prediction of the mass
function. Nevertheless, the exact analytical predictions are not
necessarily easy to compute. The often applied approximations
lead to unrealistic mass definitions and barrier thresholds. Recently \cite{MR1} described, how to implement
corrections induced by a realistic mass definition based on \cite{Bondetal}. In addition, they
implemented consistently a barrier which is not described by a
deterministic value, but scatters around its mean value \cite{MR2} -
 see also \cite{DORO,Monaco,BM,Audit,Lee1,Lee2,Desj,Casto}
for others literature on stochastic barrier. 
The same formalism had been extended in \cite{CA1,CA2,AC1,AC2}  to the case
where the mean threshold depends on the variance of the density field inspired
by the ellipsoidal collapse \cite{ST}, leading to a drifting diffusive
barrier (DDB). 
 However, the quantitative values of the barrier required to reproduce
 N-body simulations with high accuracy ($\sim 5\%$), are significantly
 different to the ellipsoidal collapse expectations. In \cite{ARSC} the
 authors showed that the discrepancy is due to the intrinsic
 assumption of the excursion set theory, which does not take into
 account that halos form at special positions in the initial matter
 density field (mostly associated with peak locations, see
 \cite{LudPorciani} \cite{BM, SMT, Roberston}) leading to a larger scatter than expected. \\ 

In this paper we investigate further the robustness of the DDB model
and its physical meaning by predicting the barrier parameters at different redshift from the
initial conditions of the N-body simulations. We also test its ability
to model the halo mass function measured from non-standard
cosmological simulations and discuss the pertinence of 
assuming non-universal parameters for the barrier.

In section \ref{EST} of this paper we first briefly review the
drifting diffusive barrier model \cite{AC1,CA1,CA2,AC2}, built upon
\cite{MR2,MR1,MR3}. Then we show that the collapse overdensity
measured in N-body simulations required to form halos is well
approximated by this model, while a different type of barrier inspired
by the physics of collapsing peaks leads to inconsistent predictions
once it is plugged into the standard excursion set
approach\cite{Bondetal}. In section \ref{spheri} we explore in detail
the large mass limit of our model.  In section 3 we use the values
describing the barrier measured from the initial conditions to predict
the halo mass function at $z=0;0.66;1;1.5$ and compare it against the results of
N-body simulations. Finally, we test the ability of the model to
describe the halo mass function in various non-$\Lambda$CDM
cosmologies in section 4.

\section{The Excursion Set Theory Barrier Modelling}\label{EST}

The first attempt to predict the halo mass function was done by Press
$\&$ Schechter \cite{PS}. The idea is to relate the number density of
halos $n(M)$ of mass $M$ to the fraction of volumes in  which
the overdensity $\delta(R)$ smoothed on scale $R(M)$ is greater than
some critical threshold.  However, the overdensity contained in those
regions can cross the threshold at different smoothing scales leading
to the so called cloud-in-cloud problem. This issue is resolved in the
excursion set theory \cite{Bondetal}, in which the smoothing scale is
decreased from infinity until the smoothed overdensity crosses the
threshold for the first time. The mass which corresponds to this scale defines the mass
of the halo. The smoothed overdensity is given by 

\begin{equation}
\delta(\textbf{x},R)=\frac{1}{(2\pi)^3}\int d^3k \,\tilde{W}(\textbf{k},R)\tilde{\delta}(\textbf{k})e^{-i\textbf{k}\textbf{x}}\label{deltas},
\end{equation}

where $\tilde{W}(k,R)$ is the Fourier transform of the filter function
which defines the mass enclosed in the region of radius $R$ with
$M=\bar{\rho} V(R)$, where $\bar{\rho}$ is the mean background matter density
and $V(R)=\int d^3x W(\textbf{x},R)$. Therefore, as the radius decreases the overdensity
$\delta$ performs a random walk and the properties of this random walk
depend on the filter function and the initial statistics of the matter
density field. Once the filter function is specified, there is a
one-to-one relationship between $R(M)$ and the associated variance of
the field 
\begin{equation}
S\equiv\langle \delta^2(R)\rangle=\sigma^2=\frac{1}{2\pi^2} \int dk \,k^2 P(k) \tilde{W^2}(k,R),
\end{equation}
where P(k) is the linear matter power spectrum. The halo mass function can be written as:

\begin{equation}
\frac{dn}{dM}=f(\sigma)\frac{\bar{\rho}}{M^2}\frac{d\log\sigma^{-1}}{d\log M}
\end{equation}

where $f(\sigma)\equiv 2\sigma^2 \mathcal{F}(\sigma^2)$ is the
so-called multiplicity function and $\mathcal{F}\equiv dF/dS$, with
F(S) the fraction of collapsed volume with mass $>M(S)$. Let us denote the collapse threshold by $B$ and $\Pi(\delta,S)$ the probability density of finding
an overdensity $\delta(R)$ smoothed on a scale R(S), which has never
crossed $B(S)$ on any larger scale. Hence, the fraction of collapsed
volume is given by  

\begin{equation}
\mathcal{F}(S)=-\frac{\partial}{\partial S} \int_{-\infty}^{B} \Pi(\delta,S)\; d\delta .
\end{equation}

Therefore, the two building blocks which need to be specified are the
criteria of collapse $B(R)$ and the PDF associated with the random walks
of $\delta(R)$, which never crossed the barrier on scales larger than
$R$. For a sharp-$k$ filter and Gaussian initial conditions, the
random walk performed by $\delta$ is Markovian, i.e. the
evolution  at a given step is independent of the previous steps. In
this particular case and for some specific barrier, the exact
first-crossing distribution can be solved fully
analytically \cite{Bondetal}. However,
the volume defined by a top hat filter  
in $k$-space leads to \cite{MR1}: $V_{sk}=6\pi^2 R^3-12\pi
R^3\int_{0}^{\infty}\cos(x) dx$. 
This volume $V(R)$ is inconsistent with the ones used for observations
and N-body simulations. There one generally assumes spherical filters with finite support. Hence, a better filter is the top hat filter in real
space, which clearly has the advantage of having a non-ambiguous
volume definition. In such a case
all the steps in a trajectory are correlated with one another
 and there is no exact analytical
solution. One way to overcome this problem is to run Monte Carlo
random walks as described in \cite{Bondetal} and solve numerically for
the first-crossing rate. However, this is computationally
expensive\footnote{See \cite{Shethnew} for a computationally cheap algorithm which returns an approximation of the first crossing distribution.}. In \cite{MR1}, the authors apply a path integral formalism
to compute analytically the non-Markovian corrections associated with
the sharp-x filter. In this framework, the problem is fully specified
by the correlators of the trajectories. For Gaussian initial conditions
the only correlators which are required to specify the statistical
properties of the density field, are the mean and the covariance. In such a case,
$\langle \delta(S) \rangle=0$ and $\langle\delta(S_1)
\delta(S_2)\rangle_c\sim\rm{min}(S_1,S_2)+\kappa S_1(S_2-S_1)/S_2$, where
$S_2>S_1$ and $\kappa$ is the amplitude of this non-Markovian
correction, which is only weakly dependent on the smoothing scale, and is set by the linear matter power spectrum. For a
$\Lambda CDM$ universe $\kappa \sim 0.45$. 
In \cite{CA2} the authors
show that the first-order correction in $\kappa$ is sufficient to
recover the exact Monte Carlo solution within $\sim 5\%$ for the model of barrier which reproduced the N-body simulations also within $5\%$. This barrier has been extended to more general shapes in
  \cite{Micki}, although in a different context. Again the agreement between theory and Monte Carlo was within $5\%$.
   In the limit of deterministic barriers the theory could require higher order
  correction in $\kappa$ \cite{MS}. However as it was shown in \cite{ARSC}, deterministic barriers cannot be used in the excursion set theory consistently with the initial threshold measured in N-body simulations. Furthermore, accurate predictions for the mass function with respect to N-body simulations require a stochastic barrier \cite{CA2}.
  
Alternative approaches have been developed in \cite{MS,PSD,FB,Lim,Shethnew}.
 

 \subsection{Mapping the Critical Threshold in N-body Simulations to the Barrier Model}
 
In the simplest version the excursion set approach assumes a
deterministic spherical collapse threshold
\cite{GunnGott1972}. However, initial overdense patches of matter are
ellipsoidal and are sensitive to both the internal and external shear
fields leading to a stochastic barrier
\cite{Alimi,ST,SMT,DES}. Furthermore, in \cite{SMT} the authors found
that on average smaller proto-halos require a higher overdensity in order to
collapse. That analysis uses the center of the proto-halos to compute
the critical overdensity. In this case the
mean is well approximated by $\langle B\rangle=\delta_c+\beta
S^\gamma$ (with $\beta\sim 0.47$, $\gamma\sim 0.615$), while its
variance is of order $\sim 0.032 S$ \cite{ARSC} (see \cite{MR2} for another derivation of the variance using the ellipsoidal collapse model). Assuming the evolution of the collapse
threshold at a given scale $R$ is uncorrelated to the collapse at
other scales, it is possible to introduce a parameter $D_B$ which
models the amplitude of the scatter around the mean of the barrier:
$\langle B(S_1)B(S_2)\rangle_c=D_B\rm{min}(S_1,S_2)$. See also \cite{ARSC} for a discussion of this model.  
For $D_B=0$, the
barrier is deterministic. One could think of estimating $D_B$ using
the ellipsoidal collapse model but this would be wrong because the
excursion set theory averages over trajectories without taking into
account whether their positions correspond to a peak or not \cite{Casto,Bondetal,MS,PSD}. 
Thus, the amplitude of the scatter based on the standard ellipsoidal
collapse is expected to be smaller compared to the variance of the
barrier at a random position inside the proto-halo. 
Indeed, our physical intuition is that once we sit on a peak, we are
less sensitive to the external shear field. Therefore the question is:
What is the barrier we need to use in the standard excursion set
framework? 
In order to account for the excursion set averaging over all
positions, \cite{ARSC} showed that one should consider a threshold
distribution which corresponds to the overdensity around a randomly
selected halo particle and not the one which corresponds to the center
of mass of the proto-halo.  
In other words, the critical overdensity distribution,
which has led to the formation of halos, can be reconstructed by
measuring the density fluctuation
$\delta(R_h)$ in the initial conditions around a randomly selected particle which 
belongs to a proto-halo of Lagrangian size\footnote{The Lagrangian radius of a halo is defined 
such that the mass inside a sphere of radius $R_h$ in the early homogeneous universe is equal to the halo mass.} $R_h$. 
By identifying the measured critical overdensity with the overdensity at first crossing of the barrier, 
i.e.~$\delta(R_h)\equiv\delta_{1x}$, one can map the measured overdensity distribution to the barrier
distribution needed for the excursion set formalism.

\begin{figure}
\begin{center}
\includegraphics[scale=0.45]{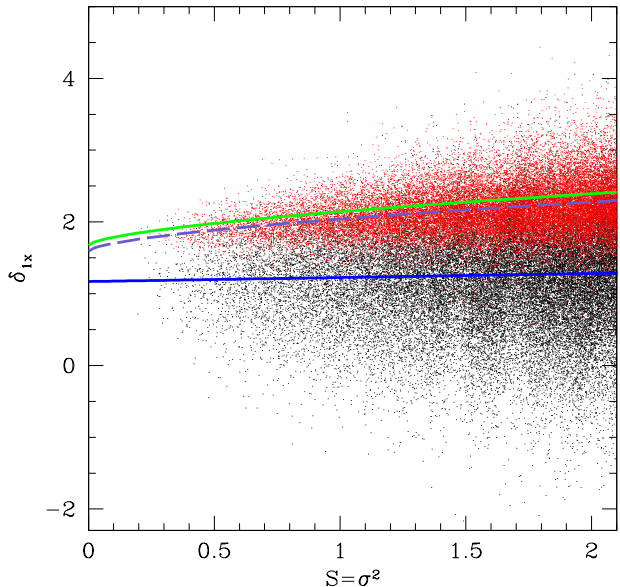}
\caption{Initial overdensities encapsulated in a the Lagrangian radius of a
  halo identified at z=0 center around randomly selected particles, 
  (black dots) and around the center of
  mass of  proto-halos (red dots). The solid green line shows the
  mean of the center of mass PDF. The blue dashed line is the
  ellipsoidal prediction\cite{ST} for the mean overdensity while the
  solid blue line is the mean prediction that we fit in
  section \ref{fitparam}. }\label{fig2} 
\end{center}
\end{figure}

To see how the critical overdensity distribution changes when we
consider a random position of the proto-halo instead of the center of
mass position, we show in Fig.~\ref{fig2} the result of the two
approaches obtained from the initial conditions of an N-body simulation. 
Red dots correspond to the
center of mass overdensity, while black dots are for randomly
selected halo particles. As expected, the scatter of the red
distribution is sharper than the black distribution. In addition,
the centre of mass overdensities are mostly positive, suggesting that most proto-halos
correspond to initial peaks \cite{LudPorciani}. The associated PDF can
be approximated by a log-normal distribution \cite{Roberston,ARSC}.
The blue dashed line in Fig.~\ref{fig2} shows the mean  prediction of
the ellipsoidal collapse \cite{ST}, while the solid green line shows
the mean that we measure for the center of mass distribution. The blue
solid line is the mean associated with the randomly selected halo
particle (black distribution). A quantitative analysis is given in the
Table 1 of \cite{ARSC}. At fixed scale $S$, it is shown in \cite{ARSC} that the
PDF of $\delta_{1x}$ at random positions inside the proto-halo is very
well approximated by a simple Gaussian distribution over the range of
masses probed. Furthermore, after mapping the barrier
distribution required to fit the N-body halo mass function to the
$\delta_{1x}$ distribution, one recovers the shape, width, and mean of
the distribution measured from the random halo particles (see
\cite{ARSC}). In contrast, in this article we start from the initial
conditions of the matter density field and infer from the proto-halos the parameters of the collapse
barrier. We then use these collapse parameters to predict the mass function. 

In order to do this we first have to establish the link between the
PDF of the barrier and the PDF of the critical overdensity threshold
$\delta_{1x}$ at which the random walk of $\delta(S)$ crosses the barrier for the first time. 
A general derivation can be found in
\cite{ARSC}, however we propose here an alternative, more straightforward
approach, which leads to the same expression as Eq.(11) in
\cite{ARSC}. \\ 

In the simple case where the barrier is given by a Gaussian with mean
$\bar{B}$  and  variance $D_B S$, the probability to obtain B on a
given scale S is  

\begin{equation}
P(B,S) \propto e^{-(B-\bar{B})^2/(2D_B S)}
\end{equation}
and for Gaussian initial distribution of the matter density field the
probability to get an overdensity $\delta$ on scale $S$ is  
\begin{equation}
P(\delta,S) \propto e^{-\delta^2/(2S)} .
\end{equation}
Therefore we can introduce the ensemble of values $\delta_{x}$ where
$B=\delta=\delta_x$. If the random walk performed by $B$ is
independent of $\delta$
 (i.e.~$\langle B(S) \delta(S)\rangle$=0 \cite{CA1,CA2,AC1,ARSC})\footnote{This assumption can be physically motivated, 
e.g.~\cite{ST} presented a model in which the collapse barrier depends on the 
traceless shear, which is independent of $\delta$.}
 Therefore, the intersection of those two sets of values $\delta_x=\left\lbrace
B\cap\delta\right\rbrace $ at a given $S$ is simply given by  
\begin{equation}
\begin{split}
P(\delta_x,S)&=P(B=\delta_x,S)P(\delta=\delta_x,S)\\
&\propto e^{-\frac{(\delta_x-\mu_{x})^2}{2 S D_{B}^{\rm{eff}}}}
e^{\frac{-\bar{B}^2}{2S(1+D_B)}}\, ,
\end{split}
\end{equation}
where
\begin{equation}
 D_{B}^{\rm{eff}}=D_B/(1+D_B) \quad {\rm and}\quad  \mu_x=\bar{B}/(1+D_B)
\end{equation} 

Thus the PDF of the crossing height $\delta_x$ is given by 

\begin{equation}
\begin{split}
\Pi(\delta_x,S)\equiv\frac{P(\delta_x,S)}{\int P(\delta_x,S)
  d\delta_x}=\frac{e^{-(\delta_x-\mu_x)^2/(2 S
    D_{B}^{\rm{eff}})}}{\sqrt{2\pi D_{B}^{\rm{eff}}S} }\,.\label{d1x} 
\end{split}
\end{equation}

To derive the PDF of $B(S)=\delta(S)$ for the first
time, i.e.~the first-crossing distribution $\delta_{1x}(S)$, we should compute the
PDF associated with $\delta_{x}(S)$ under the condition that
$\delta<B$ on all larger scale: $P(B(S)=\delta(S), S|\delta(S')<B(S')
\ {\rm for\ all } \ S'<S)$.

For Markovian walks (sharp-$k$ filter for $\delta(S)$ and $B(S)$), the above expression also describes the PDF of the \emph{first}-crossing height $\delta_{1x}$ because the walks do not keep memory of what happen at other scales.

In the general case, however, the first-crossing condition (i.e.~random walk with absorbing boundary condition) might involve non-Markovian corrections when $\delta$ is smoothed with a sharp-$x$ filter while the barrier is not \cite{Casto,ARSC}. These corrections play a role at low
masses (large $S$) and are negligible at high masses where a large collapsing patch can rarely be embedded by a larger collapsing one (cloud-in-cloud problem
\cite{Bondetal}).

 However in what follow we use an empirical approach: as we find in \cite{ARSC}, neglecting this condition only affect the mean value of $\delta_{1x}$. The agreement between the Monte Carlo and Eq.(\ref{d1x}) is quite accurate for the variance while the mean value is systematically higher for the theory compared to the Monte Carlo. In addition this shift depends on $S$. Thus we introduce a correction term $q$ such as 

\begin{equation}
<\delta_{1x}>=\frac{\delta_c+(\beta+q)S}{1+D_B}
\end{equation}
and we fixe $q$ by measuring the mean value $<\delta_{1x}>$ of the Monte Carlo. Using the values reported in the Tab 1 of \cite{ARSC}, we find in average that $q=-0.1$. Therefore, in what follows we will use Eq.~(\ref{d1x}) with 

\begin{equation}
 D_{B}^{\rm{eff}}=D_B/(1+D_B) \quad {\rm and}\quad  \mu_x=(\bar{B}-0.1 S )/(1+D_B)\label{mapping}\,.
\end{equation} 

to establish the link between the
parameters of the barrier ($\bar{B}(S)$ and $D_B$), which we use in the excursion set, and the initial critical overdensity distribution measured around
randomly selected halo particles: $\delta_{x}$.

\subsection{Measurement of the Barrier Parameters from the Intial Conditions}\label{measure}

In order to determine the values of the barrier we follow the
procedure described in \cite{ARSC}. We assume that the PDF of
$\delta_{1x}$ is Gaussian given by Eq.(\ref{d1x}) and we consider a
Gaussian barrier with free parameters $D_B$ and $\beta$ with mean
$\bar{B}=\delta_c+\beta S$ and variance $D_B S$. 
We also assume that for the mass range we probe, the linear
approximation of the barrier is accurate as suggested by
\cite{ARSC}. This assumption will be further investigated in section
\ref{spheri}.  

Using Eq.(\ref{mapping}), the two parameters $\beta$ and $D_B$ are fitted to 
the mean and variance of the initial overdensity $\delta(R_h)$ around random proto-halo particles
measured from the initial conditions of N-body simulations. 
For the analysis, we use the DEUS consortium N-body simulations \cite{DEUS}
(described in \cite{Ali,Courtin,Yann}). The box lengths of the simulations
are 162, 648 and 2592 $\textrm{h}^{-1}$ Mpc, respectively, and each simulation has
$1024^3$ particles. The simulations have been carried out using the
RAMSES code \cite{Teyssier2002}  for a $\Lambda$CDM model calibrated
to the WMAP-5-year data \cite{wmap5} (i.e.~$\Omega_m=0.26, h=0.72, \sigma_8=0.79, n_s=0.96$). The halos were identified using the
Friend-of-Friends (FoF) algorithm with linking length $b=0.2$. \\

Binning the data in $\Delta S=0.1$ and using all the mass range covered by the simulation for $S>0.5$ (i.e.~where the consistency is very accurate), we derive by a simple $\chi^2$-fit the following best-fit parameters of the DDB barrier for several redshifts:
 ($z=0,\delta_c=1.673,\beta=0.14,D_B=0.4$)  ($z=0.66,\delta_c=1.682,\beta=0.17,D_b=0.38$), ($z=1,\delta_c=1.683,\beta=0.18,D_B=0.37$), ($z=1.5,\delta_c=1.685,\beta=0.19,D_B=0.35$). We tested the covariance between $\beta$ and $D_B$ by running first a $\chi^2$ fit for $D_B$ then $\beta$ and we found consistent values with the $\chi^2$ computed on a two by two grid which set simultaneously the value of $\beta$ and $D_B$.

\begin{figure}
\begin{center}
\includegraphics[scale=0.75]{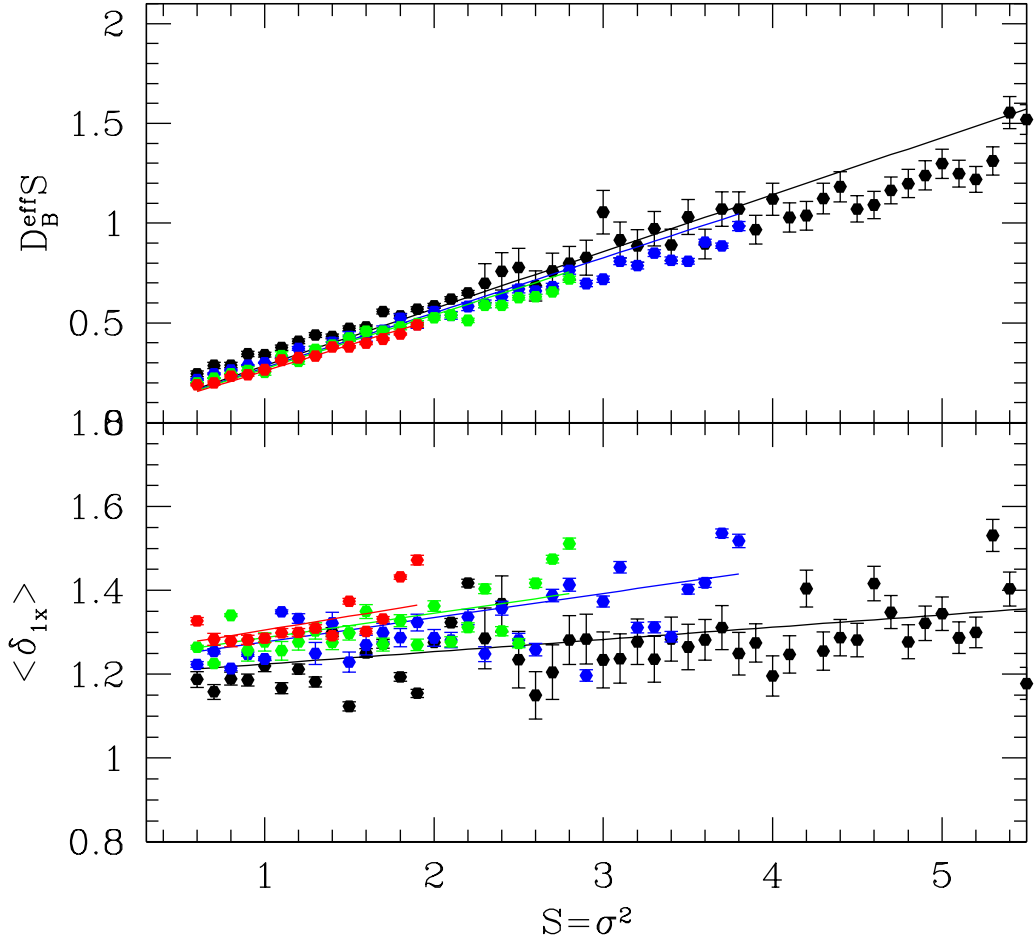}
\caption{Mean $\mu_x$ and variance $D_{\rm Beff}$ of $\delta_{1x}$ measured at redshift $z=\lbrace0,0.66,1,1.5\rbrace$ corresponding to the black, blue, green and red dots. The corresponding colored solid lines are the resulting best
  linear fit models .
   }\label{Fig6}
\end{center}
\end{figure}

 Note that these values are in good agreement with the ones found in the analysis of
 \cite{ARSC} that uses the mass function to determine those values and checks the consistency with the initial distribution of $\delta_{1x}$ over a smaller sample of mass. In addition, these values may change for different initial realizations of the density field. This cosmic variance is not taken into account in this analysis.\footnote{we thank the referee for pointing this additional source of variance.}
 Later we will use these parameters to predict the halo mass
 function. 
 In Fig.~\ref{Fig6} we show the best linear fit to the measured mean
 and variance. The error bars show the Poisson error. The rather large scatter (larger than the statistical error) in $\mu_x$ is probably due to systematics in the computation of the initial overdensity. Because particles are set on a grid at the beginning of the simulation, the estimation of the density of proto-halos is biased. Using Next Grid Point (NGP) charge assignment to estimate the density, we obtain huge spurious oscillations in $\delta_{1x}(S)$ (especially for small halos). Using Cloud In Cell (CIC) charge assignment allows us to reduce oscillations down to the level of other systematic and statistical errors.

 Overall, the linear modelling of the variance and
 the mean describes the simulation data well. However, for small $S$,
 i.e.~large halos masses, it seems that the linear model with zero variance for $S=0$ underpredicts
 the variance of the barrier. Note that standard spherical collapse predicts strictly zero scatter. This will be discussed further in the next
 section. 

Another point we can see from this figure and the predicted DDB parameters, is a consistency between their evolution on the redshift. In average $\beta$ tends to increases with z while $D_B$ tends to decreases. Since $\beta S$ is summed with $\delta_c$, it is not surprising that it should increases with z because in the spherical collapse model $\delta_c$ increases at higher redshift to reach the EdS value $\delta_c=1.686$. Perhaps more interesting is that $D_B$ seems to decreases with z. If a part of the stochasticity is due to late time merging halos then one could hope that at higher redshift the barrier is more deterministic ($D_B$ should decreases). Nevertheless the main component of this stochasticity must be attributed to averaging over random positions when we compute the multiplicity function within the excursion set theory. 

\subsection{The Large Mass Limit}\label{spheri}
A common way to compute the effect of modified gravity or dark energy
models on the formation of halos is to compute the spherical collapse
threshold in those non-standard cosmologies (e.g.~\cite{Ali,Courtin}) and use it to predict the mass function. 
As we already know, the spherical collapse is based on unrealistic assumptions: no external or internal shear.
However one could hope that in the large mass limit, where
$S\rightarrow 0$, these assumptions become more realistic. In such case,
 the mean and the variance of the
barrier should converge to the spherical collapse barrier (i.e.~$P(B,S)
\rightarrow \delta_D(B-\delta_c)$).

We test this assumption by analysing the PDF associated with
$\delta_{1x}(R_h)$ for randomly selected
halo particles and for the center of mass positions of massive halos. We measure the 
distributions from the initial conditions of the N-body simulation with a box
length of $L=2592 h^{-1} {\rm Mpc}$. This
large box enables us to probe even very massive halos with good statistics, for which one would expect the spherical collapse to be a good
approximation. Note that this large box was not used in \cite{ARSC}.

\begin{figure}
\begin{center}
\includegraphics[scale=0.45]{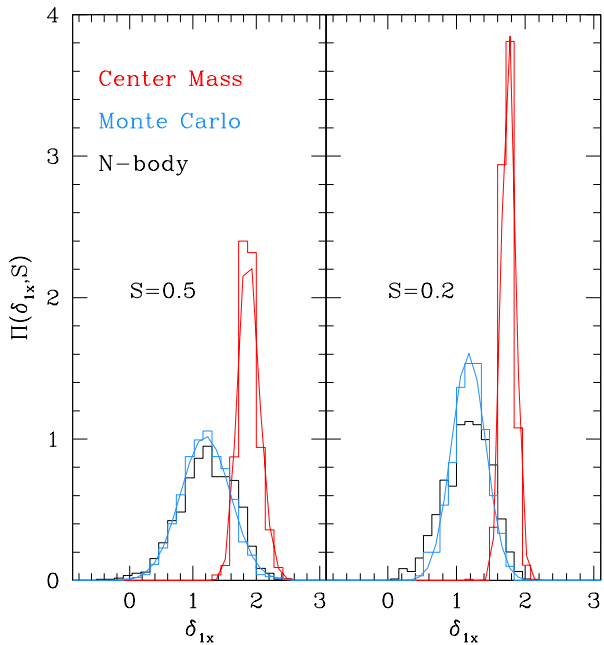}
\caption{PDF of the overdensity around center of mass (red histogram)
  and randomly selected particles (black histogram) of proto-halos of
  size $S(R)=0.5$ (left panel) and $S(R)=0.2$ (right
  panel). Blue and red solid lines correspond to Eq.(\ref{d1x}) and
  Eq.(\ref{picm}), respectively. Blue histogram correspond to the Monte Carlo PDF
  using $\beta,D_B$ which best fit Eq.(\ref{d1x})}\label{fig3} 
\end{center}
\end{figure}

Fig.~\ref{fig3} shows the distribution of the critical density contrast measured in the initial conditions - leading to the formation of halos at z=0. The two panels correspond to two different mass of halos: $S=0.2$ ($M=1.6
\times10^{15} M_{\odot}/h$) and $S=0.5\ $($M=2.6 \times10^{14} M_{\odot}/h$). 
The PDF of $\delta_{1x}$ associated with the center of mass
particles is the red histogram while for randomly selected particles it is the black
histogram. In those two bins the total number of halos are for $S=0.2$ $N=786$ and for $S=0.5$, $N=824$. The binning is in $\Delta \delta_{1x}=0.12$. 

The Monte Carlo prediction of this PDF is also shown by the blue histogram. It is obtained by generating numerically random walk with the barrier parameters measured for the $z=0$ halos from the initial conditions. We record for each walk the first-crossing values $\delta_{1x}$ (i.e.~when $\delta(S)=B(S)$) and make a histogram of those values given the two extreme masses we consider here. 
The prediction Eq.(\ref{d1x}) is the solid blue curve. For those we use the best fit values of $\beta,D_B$ that we determined in \cite{ARSC}. This figure shows that the linear barrier
modelling holds for $S=0.5$ but for very massive halos ($S=0.2$) the
width of the measured distribution is significantly larger than the
linear model predicts. The measured mean and variance are (1.16,0.11) at $S=0.2$ and (1.2,0.3) at $S=0.5$.
In addition we report a skewness $-0.01$ at $S=0.2$ and $ -0.22$ at $S=0.5$ which is compatible with the one measured in the Monte Carlo.

On the other hand, the PDF associated with the center of mass seems to converge to a deterministic value. The mean and the variance are ($S=0.5,\langle\delta_{cm}\rangle= 1.748,\langle\delta^{2}_{cm}\rangle=0.008678$)
 and ($S=0.5,\langle\delta_{cm}\rangle=1.88, \langle\delta^{2}_{cm}\rangle= 0.02539$).


As pointed out in \cite{ARSC}, the variance here is larger than that measured in \cite{Despali}. This difference might be due to the fact that we built this center-of-mass PDF assuming initial proto-halo to be spherical as it was done in \cite{Roberston}.

Using these measurements of the mean and variance and those reported in \cite{ARSC}, we find that the associate PDF can be well-fitted by the following log-normal distribution (red solid line):

\begin{equation}
\Pi(\delta_{1x,\rm {cm}},S)=\frac{e^{-\frac{(\ln[\delta_{1x,\rm cm}]-\nu_{\rm cm})^2}{2 \Sigma^2}}}{\sqrt{2\pi} \delta_{1x,\rm cm}  \Sigma}\label{picm}
\end{equation}
where $\nu_{\rm cm}=\ln[(\delta_c+\beta_{cm}S^\gamma)/q]-\Sigma^2/2$ and $\Sigma^2=aS$ with
$(a=0.016,\beta_{cm}=0.47,\gamma=0.615,q=1.05)$. Note that the mean
value is in reasonable agreement with \cite{ST}: it is systematically
rescaled by a factor $q$. On the other hand, the variance is in
disagreement with the theoretical ellipsoidal prediction from
\cite{MR2} and \cite{ARSC}. As mentioned in \cite{ARSC}, this discrepancy may be due to the
fact that our estimate of $\delta_{1x,\rm cm}$ is broadened as a consequence
of having assumed spherical proto-halo which is usually not the case. \\ 

Overall, this analysis indicates that the linear model for the mean of $\delta_{1x}$ is in
reasonable agreement with the critical overdensity measured in N-body
simulations. However, for very massive halos, the scatter around the mean is larger than expected and not consistent with the model. Hence, the predicted halo mass function might not be able to reproduce the N-body simulation in this
limit. 
This was not observed in \cite{ARSC}, where the mass function
as well as the $\delta_{1x}$ distribution was well modelled by the
linear DDB model. Note, however, that the
largest halo mass studied in \cite{ARSC} was much smaller, ($M\sim 5.2\times10^{13} M_{\odot}/h$), than in our
current study. We will test the accuracy of the mass function
predictions in Section \ref{mf}. 

One could think about solving the first-crossing distribution for a
barrier which mimics the center of mass distribution of $\delta_{1x,cm}$
(Eq.~(\ref{picm})). In the next section we will see that such a
barrier will lead to inconsistent results for the halo mass function. This confirms the idea that the standard excursion set framework
requires a barrier modelling valid for random positions and not for special positions like density peaks.

\subsection{The Barrier for Peaks}\label{Lognor}
We can perform one further test on the modelling of the barrier, which
enters the excursion set theory. Previously we assumed a Gaussian
barrier and showed that the N-body first-crossing
distribution is well reproduced by the analytical theory once we use the barrier measured from random halo particles. One might wonder
if one can also use a barrier inspired by peak statistics, which when
used in the standard excursion set framework, is able to recover the PDF of
$\delta_{1x,cm}$ associated with the center-of-mass of the proto-halo
(the red distribution in Fig.\ref{fig2} described by Eq.(\ref{picm})). In fact, we
previously saw that $\Pi(\delta_{1x,\rm cm},S)$ had a slightly
different mean and a very different variance than expected from
the ellipsoidal collapse theory (\cite{ST,MR2}). However, if we also
use a log-normal distribution for the barrier, we might expect
that the mean and the variance of the first-crossing distribution get
shifted compared to the ones we set for $B$ (See also \cite{Casto}). 
Thus we might ask: what is
the PDF of the barrier which at first crossing leads to Eq.(\ref{picm})? And
subsequently, is a log-normal barrier able to predict the halo mass
function measured from N-body simulations?\\ 

To answer the first question, we ran Monte Carlo walks using a PDF for
the barrier calibrated on the center-of-mass PDF Eq.(\ref{picm}) (with
$q=1.05$). In fact it is easy to extend the generation of random walks for barriers which have a log-normal distribution instead of a normal one. If $B$ is a random variable which has a normal distribution with mean $\nu$ and variance $\sigma$,
then $B_{cm}=\exp(B)$ will have a log-normal distribution with a mean value $e^{\nu+\sigma^2/2}$ and variance $(e^{\sigma^2}-1)e^{2\nu+\sigma^2}$. Inverting these relations, we generated Monte Carlo walks using a log-normal distribution for the barrier. 

It is easy to check the resulting Monte Carlo distribution is in good agreement with the theoretical formula of a log-normal distribution. 

We found that at first crossing the PDF of $\delta_{1x,\rm
  {cm}}$ is also a log-normal with a mean shifted to lower values,
which can be fitted by a factor $q=1.1$. Thus in order to reproduce
the PDF of $\delta_{1x,\rm cm}$ that we measured from the initial
conditions of the N-body simulations, we should use a log-normal
barrier in the excursion set as given by Eq.(\ref{picm}) with
parameters $\sim(a=0.016,\beta_{cm}=0.47,\gamma=0.615,q=1)$. However,
the associated mass function is completely offset to the mass function
measured from N-body simulations. This is not surprising since the
excursion set is unable to distinguish whether the trajectories belong
to the center of the peak or not. The PDF of first-crossing returned
by the excursion set theory is the one of $\delta_{1x}$ for a randomly
selected particle. Therefore any barrier which predicts only positive
values of $\delta_{1x}$ (such as the log-normal) cannot lead to a
consistent distribution of $\delta_{1x}$, especially at low masses
where, as shown in Fig.~\ref{fig2}, a non-negligible part of
$\delta_{1x}$ is negative.\\ 
 Nevertheless, one could argue that the consistency is not important
 as long as we can use a PDF of the barrier predicted by theory which
 can reproduce the mass function. Currently, however, there is no
 theory able to do so. For instance, the ellipsoidal collapse
 prediction of the barrier cannot reproduce the halo mass function
 once it is used in the \emph{standard} excursion set approach as
 we investigate here or in the excursion set peak formalism as
 studied by \cite{PSD}. In fact in \cite{PSD} the authors had to calibrate their
 barrier model based on the analysis of \cite{Roberston}. Thus the
 main advantage of this drifting diffusive barrier is that the only
 two parameters which described the barrier can be measured in the
 initial conditions and lead to a very good agreement with respect to
 the mass function as we show in the next section.

\section{The Halo Mass Function}\label{mf}
\subsection{The Drifting Diffusive Barrier Halo Mass Function}
For the DDB model barrier that we reviewed in the previous section,
the corresponding multiplicity function for a generic initial statistics
described by the second and third-order correlation functions is given
by \cite{CA2,AC1,AC2} 
\begin{equation}
f(\sigma) = f_0(\sigma) + f_{1,\beta=0}^{m-m}(\sigma)
          + f_{1,\beta^{(1)}}^{m-m}(\sigma) +
          f_{1,\beta^{(2)}}^{m-m}(\sigma) +
          f_{NG}(\sigma)+o(\kappa^2)\label{hmf}\, ,
\end{equation}
with
\begin{equation}
  f_0(\sigma)=\frac{\delta_c}{\sigma}\sqrt{\frac{2a}{\pi}}\,e^{-\frac{a}{2\sigma^2}(\delta_c+\beta\sigma^2)^2}\, ,
\end{equation}
\begin{equation}
f_{1,\beta=0}^{m-m}(\sigma)=-\tilde{\kappa}\dfrac{\delta_c}{\sigma}\sqrt{\frac{2a}{\pi}}\left[e^{-\frac{a \delta_c^2}{2\sigma^2}}-\frac{1}{2} \Gamma\left(0,\frac{a\delta_C^2}{2\sigma^2}\right)\right],
\end{equation}
\begin{equation}
f_{1,\beta^{(1)}}^{m-m}(\sigma)=- a\,\delta_c\,\beta\left[\tilde{\kappa}\,\text{Erfc}\left( \delta_c\sqrt{\frac{a}{2\sigma^2}}\right)+ f_{1,\beta=0}^{m-m}(\sigma)\right],
\end{equation}
\begin{equation}\label{beta2}
f_{1,\beta^{(2)}}^{m-m}(\sigma)=-a\,\beta\left[\frac{\beta}{2} \sigma^2 f_{1,\beta=0}^{m-m}(\sigma)+\delta_c \,f_{1,\beta^{(1)}}^{m-m}(\sigma)\right],
\end{equation}

where $\tilde{\kappa}=a\,\kappa = \kappa/(1+D_B)$ and $\kappa=0.465$ for a vanilla $\Lambda$CDM universe\footnote{the value of $\kappa$ is set by the linear matter power spectrum so its value does not change much for realistic cosmological scenarios} and the primordial non-Gaussian term is
\begin{equation}\label{fngl}
\begin{split}
f_{NG}(\sigma)&=\frac{a}{6}\sqrt{\frac{2a}{\pi}}\sigma e^{-\dfrac{a(\delta_c+\beta\sigma^2)^2}{2\sigma^2}}\times\\
&\biggl\{S_3(\sigma)\biggl[\dfrac{a^2}{\sigma^4}\delta_c^4-2\dfrac{a}{\sigma^2}\delta_c^2-1+3\dfrac{a^2}{\sigma^2}\beta\delta_c^3+3 a\beta \delta_c+a^2\beta^3\sigma^2 \delta_c+3 a^2\beta^2\delta_c^2+13 a \beta^2\sigma^2\biggr]+\\
&+\dfrac{dS_3(\sigma)}{d\log\sigma}\biggl[\dfrac{a}{\sigma^2}\delta_c^2-1+3 a\beta \delta_c+4 a\beta^2\sigma^2\biggr] \biggr\}+\\
&\dfrac{2}{3}a^3\beta^3\sigma^4 e^{-2a\beta \delta_c}\text{Erfc}\biggl[\sqrt{\dfrac{a}{2\sigma^2}}(\delta_c-\beta\sigma^2)\biggr]\biggl\{ 4 S_3(\sigma)+\dfrac{dS_3(\sigma)}{d\log\sigma}\biggr\} 
\end{split}
\end{equation}

where the skewness $\mathcal{S}_3(\sigma)\equiv \langle \delta^3(\sigma)\rangle/\sigma^4$ is given by Eq.(15) in \cite{AC1} and depends on the inflationary scenario. For an extension to the trispectrum see \cite{AC2}.

\subsection{Comparison with N-body Simulation Using the Physical Barrier Parameters}\label{fitparam}
\begin{figure}
\begin{center}
\includegraphics[scale=0.45]{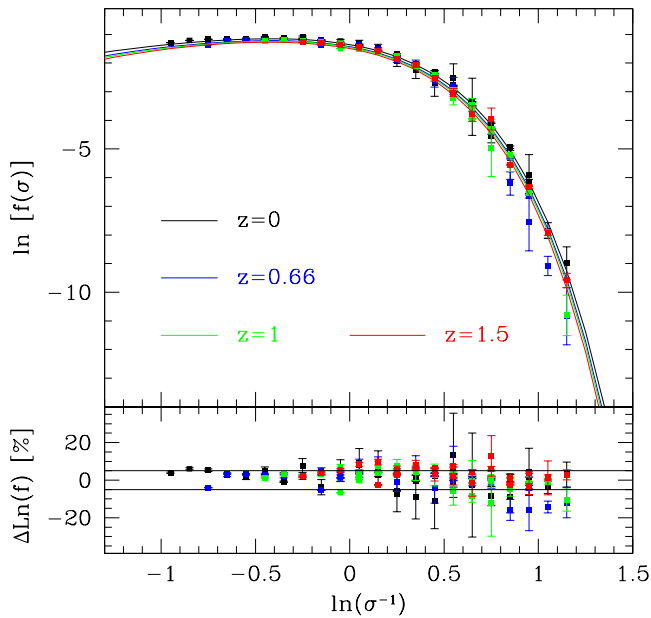}
\caption{(Upper panel) Multiplicity $f(\sigma)$ of the DDB model as given by
  Eq.~(\ref{hmf}) for the Gaussian $\Lambda$CDM case at different redshift  (solid
  lines) for $\beta, D_B$ measured in the initial conditions. The data points
  show the multiplicity function of FoF(b=0.2) halos from N-body
  simulations at the corresponding redshift. Lower panel shows the relative difference with respect to
  Eq.~(\ref{hmf}). The thin black solid lines indicates $5\%$
  deviations.}\label{fig1} 
\end{center}
\end{figure}

In this section we test the DDB prediction for the mass function in
the Gaussian $\Lambda$CDM case using the parameter $\beta$ and $D_B$
that we measured from the initial conditions (see Section
\ref{measure}). This is performed by using the publicly available halo catalogues
of the simulations we obtained from the DEUS collaboration,
where halos had been 
identified using the FoF algorithm with a linking length of $b=0.2$. Results
are shown in Fig.~\ref{fig1}, dots depict the N-body simulation
data and the solid lines show the corresponding theoretical predictions Eq.(\ref{hmf}) using the collapse parameters we measured in the initial conditions (see section 2). Different color show the different redshift.  

The agreement with respect to the N-body simulation is at $\sim 5\%$
level, which is impressive since our model for the barrier is rather
simple and we use a simple mapping with Eq.~(\ref{mapping}) to predict the
parameters for collapse. In fact, the agreement is less accurate at low
mass compared to \cite{ARSC} where the mapping is expected to be less
accurate. 

\subsection{Redshift evolution and non-universality of the collapse barrier}
Cosmic structures carry a fossil record of the past cosmic evolution \cite{Ali}. The multiplicity function therefore depends on the properties and nature of dark energy which governs the expansion rate \cite{Courtin}. Within the $\Lambda$CDM cosmology, it has been shown that the multiplicity function increases ($\sim 10\%$) at low redshift because the dark energy density $\Omega_{DE}(z)$ increases. At high redshift it converges to Einstein-de-Sitter values since $\Omega_{DE}(z)\rightarrow 0$. By looking very carefully at Fig.~\ref{fig1}, we confirm this trend. The multiplicity at z=0 is slightly above the one at higher redshift and the multiplicity function at z=0.6, 1 and 1.5 seems too converge: this was already shown in great details in \cite{Courtin}. In Fig.~\ref{Fig6}, we show the mean and variance of the proto-halo overdensity $\delta_{1x}$ in the initial density field. We could clearly see the decrease of the mean and the increase of the variance of $\delta_{1x}$ at low redshift (black dots). This is consistent with the trend of the mass function since decreasing $\beta$ (and $\delta_c$) as well as increasing $D_B$ tend to increase the mass function (see for instance, \cite{CA2}). At higher redshift, we also find the same order (z=0.6 then z=1, finally z=1.5) as well as a convergence towards self-similar values. By using these parameters $\beta$ and $D_B$ (which depend on redshift) within the excursion set framework, we self-consistently recover the measured multiplicity function and its variation with redshift (Fig.~\ref{fig1}). It means that we are able for the first time to rewrite the problem of the non-universality of the mass function, as a problem of non-universality of the collapse barrier. It opens a new way to deal with the non-universality of the mass function. The original approach followed by \cite{Courtin,More2011} is to adjust the halo-finder detection threshold as a function of $\Delta_{\textrm vir}$ so that the multiplicity function is always the same. The approach suggested by this paper would be to predict or tabulate the variation of the barrier parameters as a function of the dark energy model. Fig.\ref{Fig6} already shows that the deviations of $\beta$ and $D_B$ are correlated to $\Omega_{DE}(z)$ in $\Lambda$CDM. Our goal here was to further demonstrate the self-consistency of Excursion Set Theory when varying the amount of dark energy:
we can predict the halo mass function using a barrier which is set in the
initial conditions of the N-body simulations. Currently, there is no
successful theory able to predict this barrier from first
principles. Note, however, that also for the critical overdensity
around the center-of-mass of the proto-halo, the standard ellipsoidal
collapse \cite{ST} is unable to accurately predict the variance of the
distribution (see Table 1 in \cite{ARSC}) as we mentioned already in
section \ref{Lognor}.

\section{Extension to Non-Standard Cosmology}
In this section we study if the DDB model is also able to describe the
halo mass function in non-standard cosmologies. First, we consider a
model with different dynamics than $\Lambda$CDM, namely a model with
dynamical dark energy. Then, we alter the initial
conditions and look at models with primordial non-Gaussianity.

\subsection{Quintessence Model of Dark Energy}

\begin{figure}
\begin{center}
\includegraphics[scale=0.45]{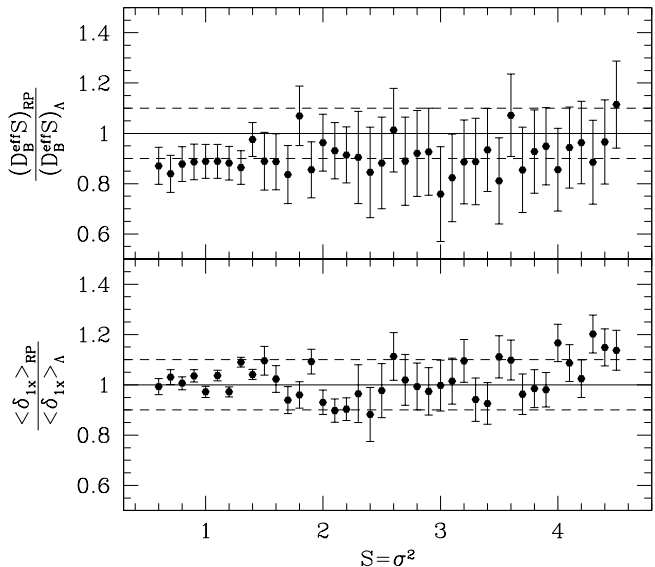}
\caption{ The ratio between the mean and the variance of Eq.(\ref{d1x}) measured in N-body simulations for the RPCDM model and the $\Lambda CDM$. Dashed line show $10\%$ deviation.}\label{fig4bis} 
\end{center}
\end{figure}

\begin{figure}
\begin{center}
\includegraphics[scale=0.45]{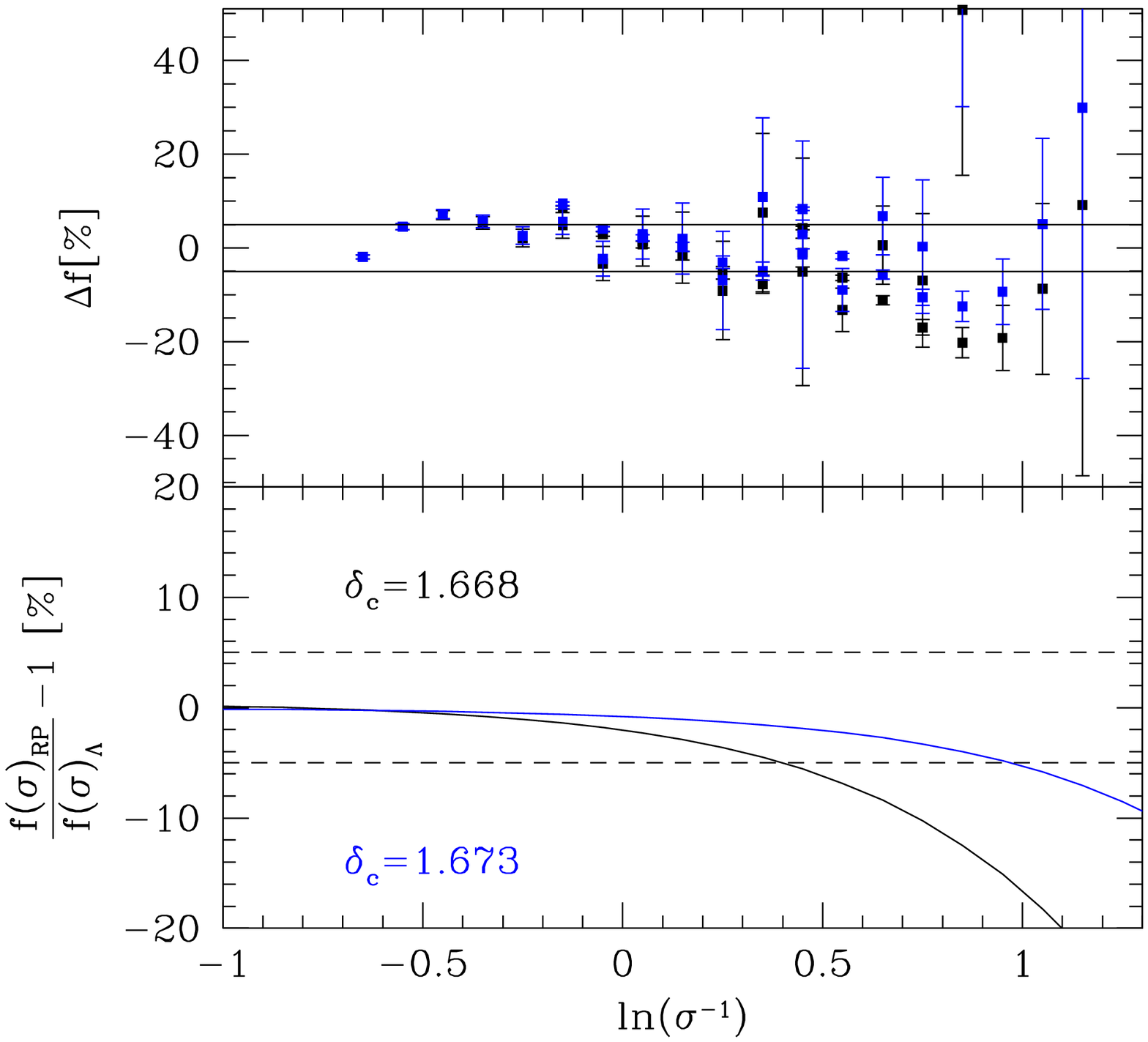}
\caption{Upper panel: relative difference between the multiplicity function measured in N-body simulation for RPCDM and the prediction of Eq.(\ref{hmf}) using the best
  fitting values of $\beta,D_B$ fixing $\delta_c$ to the RPCDM case (black dots) and fixing $\delta_c$ to the $\Lambda$CDM case (blue dot). 
  Bottom panel: relative difference between the theoretical multiplicity function of the RPCDM model with respect to the $\Lambda$CDM. Blue line is for $\delta_c=1.673$ while black line is for $\delta_c=1.668$. }\label{fig5} 
\end{center}
\end{figure}

We test the robustness of the DDB model against a change in the
dynamics of the universe. In particular, we consider a Ratra-Peebles
(RPCDM) model of quintessence \cite{RP}, which is a cosmological model
with a different background expansion to the standard $\Lambda$CDM
case \cite{Courtin,Ali,Yann} and cosmological parameters $\Omega_m=0.23,h=0.72,n_s=0.96$ in a flat universe.
 The N-body simulation data of this model
is available on \cite{DEUS}. In this model the dark energy has a
potential given by \cite{RP} 
\begin{equation}
V(\phi)=\frac{\lambda^{4+\alpha}}{m_{Pl}^{\alpha}\phi^{\alpha}}
\end{equation}
where $\phi$ is the quintessence field and $\alpha$, $\lambda$ are
respectively the slope and the amplitude of the scalar
self-interaction. The N-body simulation of \cite{Ali,DEUS} for
$RPCDM$ used $\alpha=0.5$. The dark energy state parameters are $w_0=-0.87,w_1=0.08$ and the linear
matter density fluctuations such that at $z=0$ the normalization is
$\sigma_8=0.66$. Those choice of parameters are motivated by the analysis of WMAP5 and the UNION SN \cite{wmap5,SNpaper} such that this model is not ruled out by those observations. Note that the realization of the initial Gaussian random field is identical to the $\Lambda$CDM simulations we presented above.  \\

 First, we set the barrier for $S=0$ to $\delta_c=1.668$,
which corresponds to the spherical collapse value for this particular
cosmological model \cite{Courtin}.   
Note, however, that the halo catalogue obtained from the simulation consists of FoF halos with a linking
length of 0.2 times the mean particle separation. While for a $\Lambda$CDM cosmology the FoF halo mass roughly corresponds to the virial halo mass at $z=0$ \cite{More2011}, this is not the case for the RPCDM model. In this cosmological model, the virial halo mass is actually smaller than the FoF halo mass. Hence, the obtained $\beta$ and $D_B$ might be biased and we do not expect that they  would match with the collapse parameters measured from the initial conditions. In fact it could be that requiring a different halo finder than $b=0.2$ would for instance remove low masses halos. In this case the halos used to reconstruct the barrier would not probe the high S values of the initial distribution of the barrier.

When analysing observational data, however, one does not know the true cosmological model a priori and usually would use $\Lambda$CDM as the fiducial cosmological model. We therefore want to test if the DDB model can still describe the halo mass function of RPCDM when assuming the wrong value for $\delta_c$.

In principle one could infer the correct value of $\delta_c$ by reconstructing the PDF Eq.(\ref{d1x}) in the limit $S=0$. However we would be limited by the range cover in the simulation. Nevertheless one could test if the other parameters $\beta,D_B$ differ from the $\Lambda$CDM ones. Therefore we have measured the mean and the variance of Eq.(\ref{d1x}) in the initial condition and in Fig.~\ref{fig4bis} we show the ratio with respect the ones measured in the $\Lambda$CDM simulation. First of all note that for the spherical collapse criteria, this ratio is of order $~0.997$ which would be almost impossible to extract from the numerical uncertainty. However as we can see in Fig~(\ref{fig4bis}), the variance of the $RPCDM$ PDF tends to be lower than the $\Lambda$CDM while the mean values does not differ significantly. Thus to predict the mass function of the RPCDM model, we first set the value of $\delta_c=1.668$ and using Eq.(\ref{mapping}) we predict $\beta=0.14,D_B=0.39$. Then we use instead $\delta_c=1.673$ to see if this would change the inferred value of $\beta,D_B$. However using Eq.(\ref{mapping}), we find the inferred values of $\beta,D_B$ does not changed. Note that the range of mass covered by the RP model is smaller than  the $\Lambda CDM$ simulation. This could affect the determination of the parameters since the distribution is less constrained at low masses ($S>4$).

In the upper panel of Fig.(\ref{fig5}), we can see the relative difference for the multiplicity function using the RPCDM N-body simulation (upper panel). The black dots show the prediction using the set of parameters ($\delta_c=1.668,\beta=0.14,D_B=0.39$) while the blue dots correspond to ($\delta_c=1.673,\beta=0.14,D_B=0.39$). The change in $\delta_c$ results in a small difference which affect only the high mass tail of the mass function. Overall this plot shows that both predictions are in good agreements with the simulations even if it seems the second set of parameters are in better agreement. On the lower panel, we show the relative difference between the theory of the RPCDM mass function with respect to the $\Lambda$CDM one. For the later we use the same parameters we measured in the previous section ($\delta_c=1.673,\beta=0.14,D_B=0.4$) while for the RPCDM we use the two sets of parameters ($\delta_c=1.668,\beta=0.14,D_B=0.39$ and ($\delta_c=1.673,\beta=0.14,D_B=0.39$). As we can see, the difference between the two becomes large if one use the predicted RPCDM value (black line). \\

It has been shown in \cite{Courtin} that the deviation to universality is higher if the algorithm finder used to detect halos is adjusted according to the virial density criteria of the considered cosmology. Similarly one could expect the same for the linear spherical threshold since there is a one-to-one mapping between the two. Hence, it would be interesting to test the sensitivity of the linear barrier using different halos finder criteria. We leave it for a future work.




 Finally, this analysis shows that once again the
DDB model can describe the mass function with high accuracy using only
two parameters while fitting functions of the halo mass function calibrated with N-body simulations depend on several parameters. 
Indeed, in \cite{CA2}, the authors showed that the
DDB model can reproduce the simulation fits over the entire redshift range used
in the analysis of \cite{Tinker2008} and for different redshift. In addition, the number of
parameters usually increases further when the initial conditions are
not Gaussian (e.g.~\cite{PPH8}). In the next section we will consider
different types of primordial non-Gaussianities and test if the DDB
model is still able to reproduce the N-body simulations data without introducing new
parameters.

\subsection{Primordial Non-Gaussianity}
Primordial non-Gaussianity (PNG) is generated in many inflationary
models (see \cite{NG_review} for a review), and hence measuring the
type and amount of PNG provides a unique window into the physics of
inflation. At leading order, non-Gaussianity can be described by the
bispectrum, which is the Fourier transform of the 3-point correlation
function. The amplitude of the primordial bispectrum is parametrized
by $f_{NL}$ and its dependence on the triangle configuration (formed
by the three wave numbers $k_1$, $k_2$, and $k_3$) is called shape
dependence. Current constraints from observations of the cosmic
microwave background (CMB) \cite{Planck_NG} and the large-scale
structure of the universe \cite{tom} on the level of PNG for different
shapes are all consistent with zero, i.e. no significant deviations
from primordial Gaussianity has been observed. 

As the halo mass function is sensitive to the initial conditions, it
is also a probe of PNG \cite{MVJ}. Compared to the CMB, which probes mostly large scales, the halo mass
function measures the PNG on smaller scales (roughly on the scale of
the Lagrangian radius of the halo, e.g.~a few Mpc). In contrast to the scale-dependent
halo bias induced by PNG \cite{dalal, MV}, the halo mass function is
sensitive to basically all shapes of PNG \cite{LV, wagner1} and not
only to shapes which diverge in the squeezed limit (e.g.~the local
type). This makes the halo mass function a
complimentary probe of PNG.

\begin{figure}
\begin{center}
\includegraphics[scale=0.7]{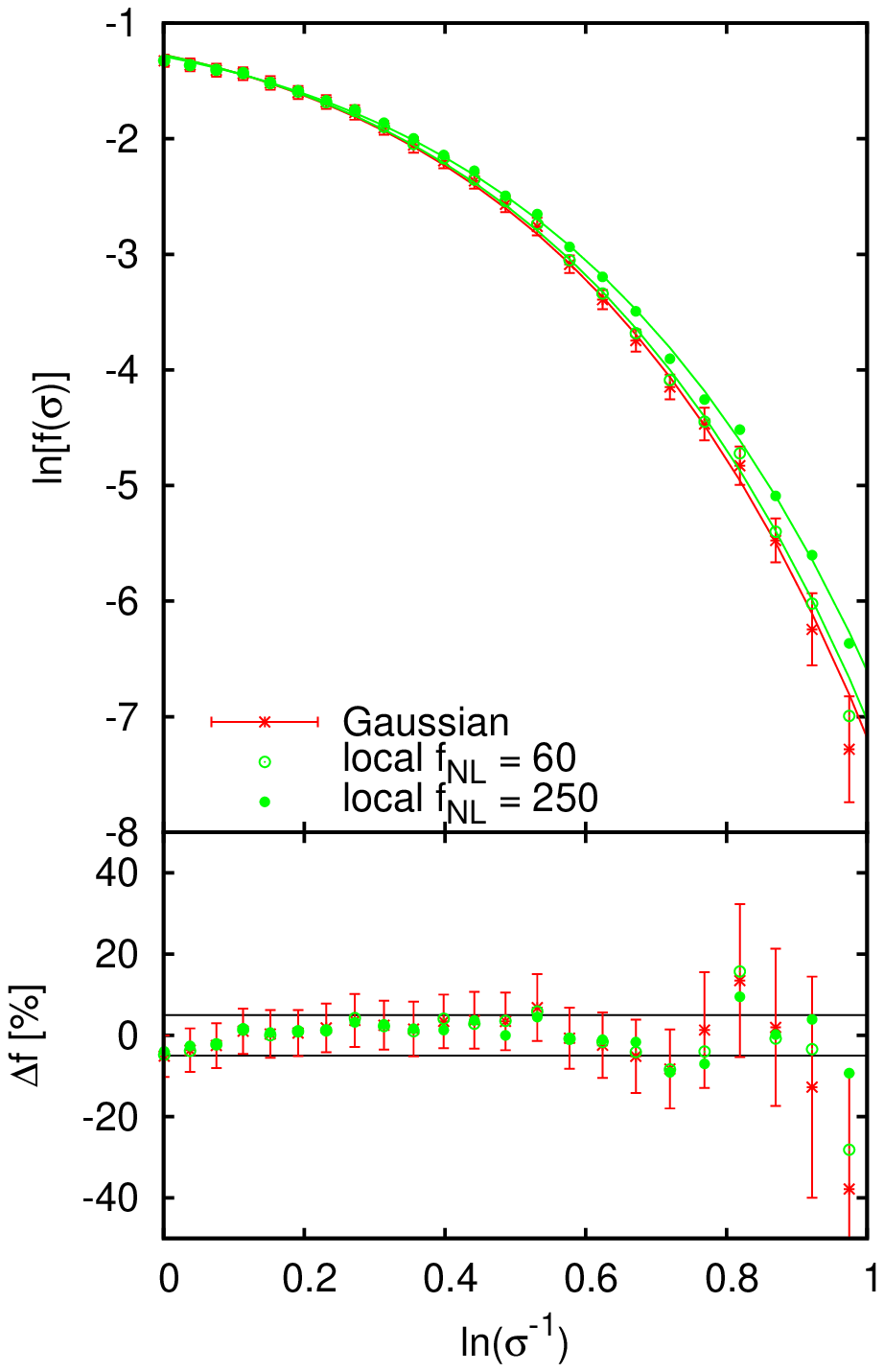}
\includegraphics[scale=0.7]{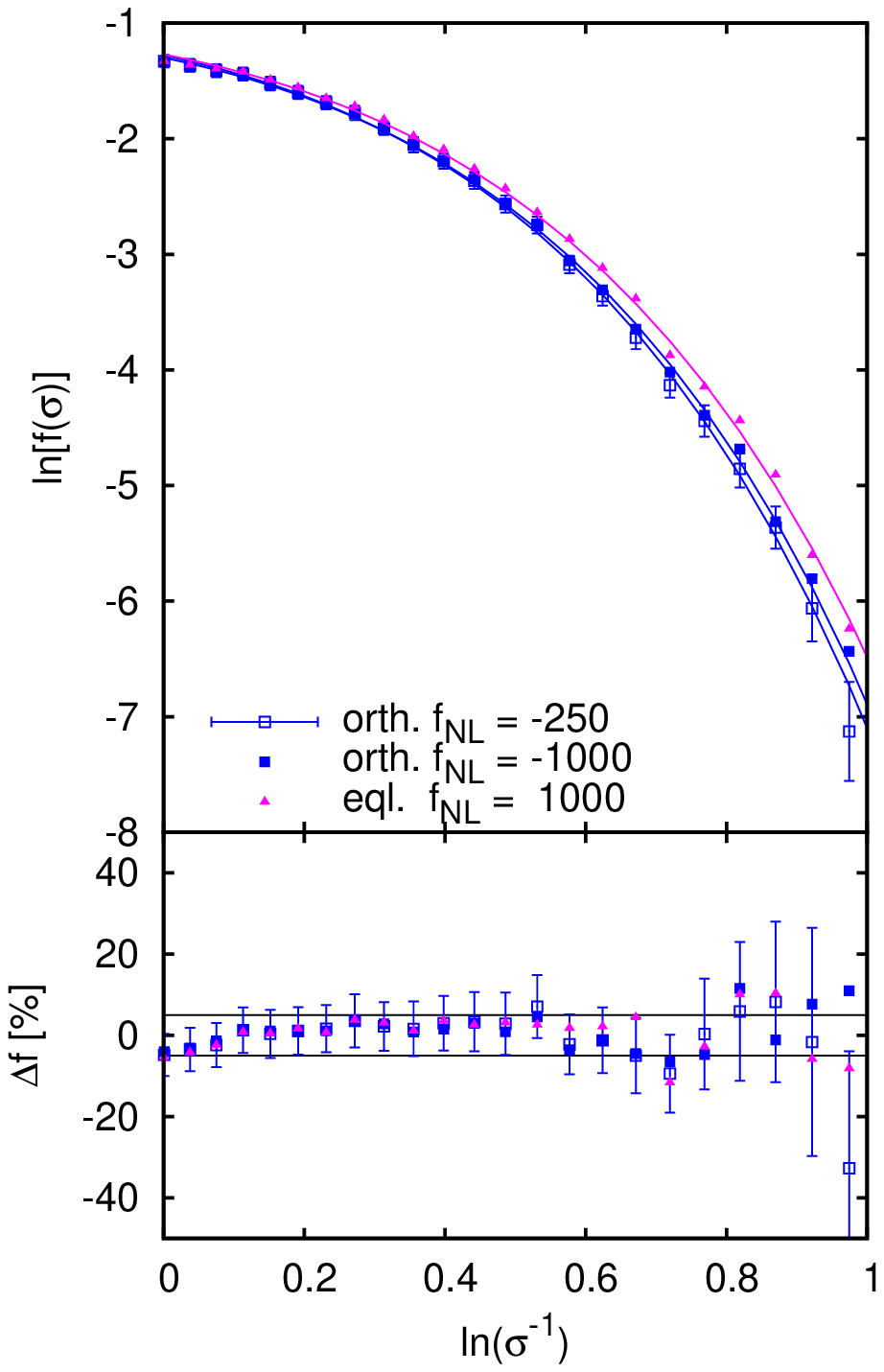}
\caption{Multiplicity functions $f(\sigma)$ measured from N-body
  simulations (data points) and the corresponding best-fit DDB model
  predictions (solid lines) at $z=0$ for various types of primordial
  (non)-Gaussianity. Lower panels show the residuals in percent
  together with +/- 5\% lines. For clarity, we only show the error bars of one model in each panel. The error bars of the other models have virtually the same size.}\label{fig_ng_mf} 
\end{center}
\end{figure}

In this section, we show that the DDB model using Eq.~(\ref{hmf}) and
Eq.~(\ref{fngl}) is able to model the halo mass function of different
types of PNG with high accuracy and without introducing a new fudge factor. The only parameters which we allow to vary are the two collapse parameters $D_B$ and $\beta$, which we fit to the mass
function derived from N-body simulations. Using the simulations of
\cite{wagner2}, we use the halo finder of \citet{AHF} to generate SOD halo catalogues
with halo masses $M_{200}$ defined by the mass inside a sphere with
mean overdensity of 200 times the mean background matter density. \\

The simulations were run with Gadget-2 and assumed a flat $\Lambda$CDM cosmology with $\Omega_m=0.27$, $h=0.7$, $n_s=0.95$, and $\sigma_8=0.7913$. The box size is 1875 $h^{-1}{\rm Mpc}$ and $1024^3$  particles were used to sample the matter distribution. Further details of the
simulations and the generation of the non-Gaussian initial conditions
as well as a description of the different simulated types of PNG can
be found in \cite{wagner2}. 
Here, we consider three different shapes of PNG: local with
$f_{NL}=60$ and $250$, equilateral with $f_{NL}=1000$, and orthogonal
with $f_{NL}=-250$ and $-1000$. Although such large values of $f_{NL}$
are already ruled out by current observational constraints
\cite{Planck_NG}, they are still useful for model testing. 
The three different types of PNG differ strongly in the shape of their bispectrum: The bispectrum of the local type diverges for squeezed triangles whereas the bispectrum of the equilateral type peaks for equilateral triangles. The bispectrum shape of the orthogonal type is mathematically orthogonal to the equilateral and local shape. However, for the effect of PNG on the mass function (see Eq.~\ref{fngl}), the most relevant quantity is the skewness $\mathcal{S}_3$, which is given by an integral over the bispectrum. Hence, the mass function is not very sensitive to the specific shape of the bispectrum but rather to the overall level of PNG.

In Fig.~\ref{fig_ng_mf}, we show the measured multiplicity functions
together with the best-fit DDB model predictions at $z=0$. From the lower
panels, which show the residuals in percent, it becomes apparent that
the DDB model is capable of describing the simulation results within
$\sim 5\%$ accuracy.

\begin{figure}
\begin{center}
\includegraphics[scale=0.4,angle=270]{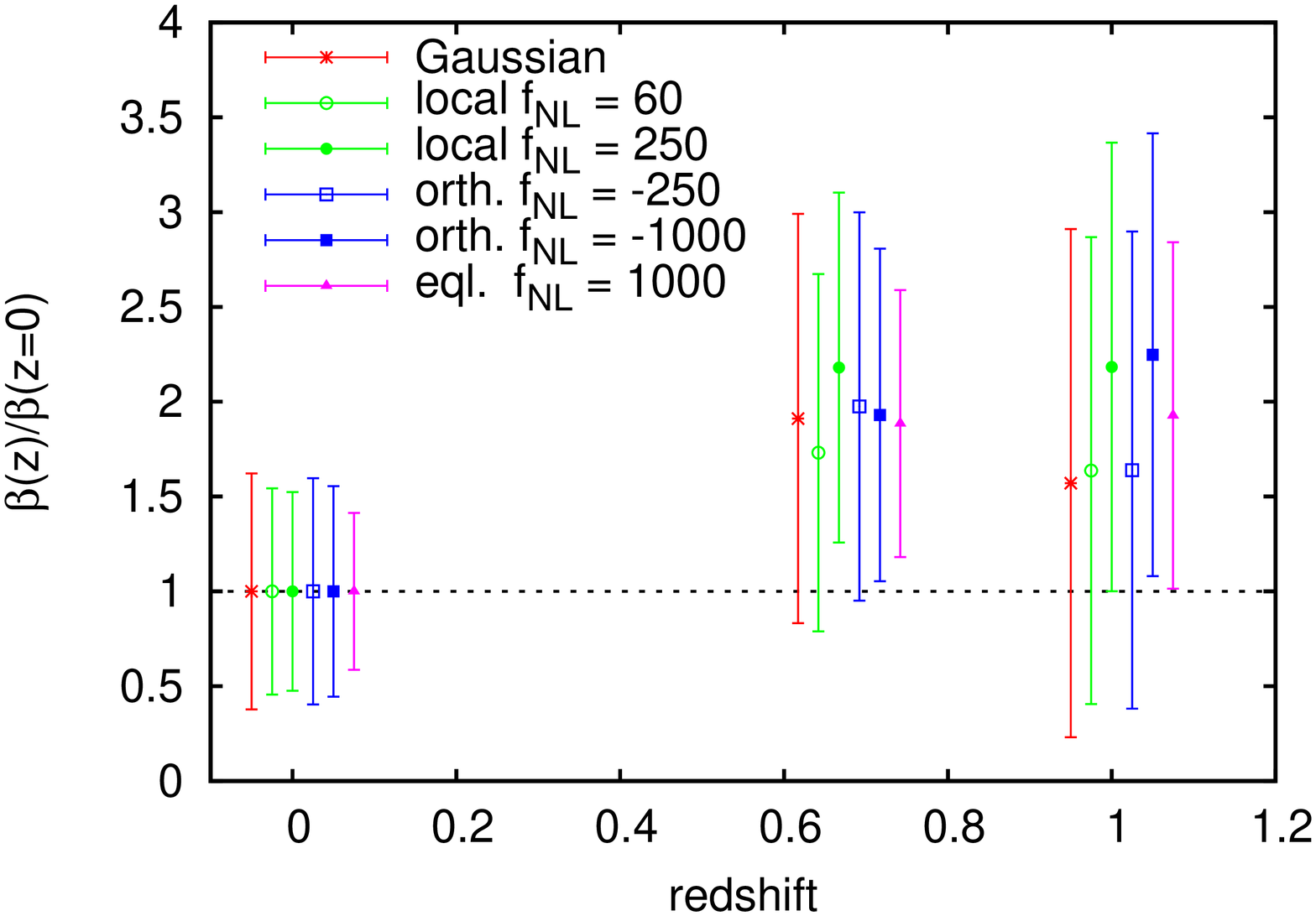}
\includegraphics[scale=0.4,angle=270]{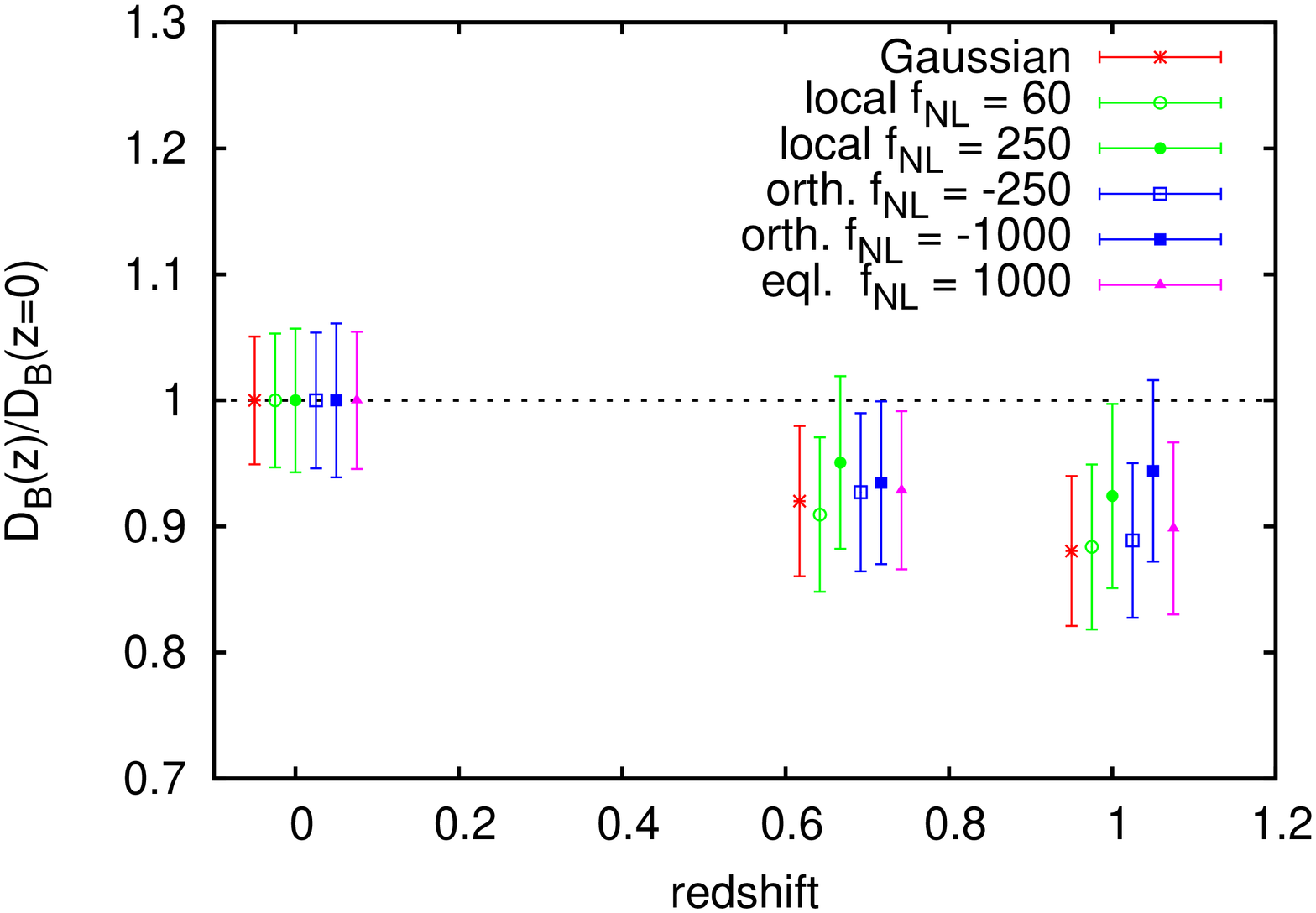}
\caption{Ratio of the best fit parameters $\beta,D_B$ at different
  redshift with respect to $z=0$ for different types of PNG.}\label{Fig7}
\end{center}
\end{figure}

\begin{ctable}
  [caption={Best-fit values of the barrier parameters for different PNG models and at different redshifts. Note that due to the small mass range probed by the simulations $D_B$ and $\beta$ are strongly correlated (their correlation coefficient is $\sim 0.95$).}, label={table}]{cccc}{}
{\FL
PNG & redshift & $D_B$ &  $\beta$\ML
Gaussian & 0 & $0.302 \pm 0.015$ &  $0.032 \pm 0.020$\NN
Gaussian & 0.67 & $0.278 \pm 0.018$ & $0.060 \pm 0.034$  \NN
Gaussian & 1 & $0.266 \pm 0.018$ &  $0.050 \pm 0.042$ \ML
local $f_{NL}=60$ & 0 & $0.310 \pm 0.016$ & $0.038 \pm 0.021$ \NN
local $f_{NL}=60$ & 0.67 & $0.282 \pm 0.019$ &  $0.065 \pm 0.036$ \NN
local $f_{NL}=60$ & 1 &  $0.274 \pm 0.020$ & $0.062 \pm 0.046$\ML
local $f_{NL}=250$ & 0 & $0.320 \pm 0.018$ & $0.042\pm 0.022$ \NN
local $f_{NL}=250$ & 0.67 & $0.304 \pm 0.022$ & $0.093 \pm 0.039$ \NN
local $f_{NL}=250$ & 1 & $0.296 \pm 0.023$ & $0.093 \pm 0.050$ \ML
orth. $f_{NL}=-250$ & 0 & $0.300 \pm 0.016$ & $0.034 \pm 0.020$\NN
orth. $f_{NL}=-250$ & 0.67 & $0.278 \pm 0.019$ & $0.067 \pm 0.035$  \NN
orth. $f_{NL}=-250$ & 1 & $0.266 \pm 0.018$ & $0.056 \pm 0.043$\ML
orth. $f_{NL}=-1000$ & 0 & $0.291 \pm 0.018$ & $0.040 \pm 0.022$ \NN
orth. $f_{NL}=-1000$ & 0.67 & $0.272 \pm 0.019$ & $0.076 \pm 0.035$ \NN
orth. $f_{NL}=-1000$ & 1 & $0.275 \pm 0.021$ &  $0.089 \pm 0.046$ \ML
eql. $f_{NL}=1000$ & 0 & $0.350 \pm 0.019$ &  $0.054 \pm 0.022$ \NN
eql. $f_{NL}=1000$ & 0.67 & $0.325 \pm 0.022$ & $0.102 \pm 0.038$ \NN
eql. $f_{NL}=1000$ & 1 &  $0.314 \pm 0.024$ &  $0.105 \pm 0.050$ \LL
}
\end{ctable}

We summarize the best-fit values of $D_B$ and $\beta$ for all models at three different redshifts (0, 0.67, and 1) in Table~\ref{table}. These best-fit values are sensitive to the range of mass covered by the N-body mass function and its error bars.  
As in the Gaussian case, the best-fit values of $D_B$ and $\beta$ vary
with redshift, as shown explicitly in Fig.~\ref{Fig7}, and halo mass definition \cite{CA2} (not shown here). 
Within the errors, the evolution with redshift of $D_B$ and $\beta$ appears
independent of the nature of the initial conditions as can be seen in
Fig.~\ref{Fig7}. 

Hence, the simulation data suggest a single redshift evolution for all models.
Note that the simulation data at different redshifts is based on the same realization of the initial density field, therefore cosmic variance, which would be present in observational data, is cancelled to a large extent in Fig.~\ref{Fig7}.  This is a concern for comparisons to observations. However, when trying to understand the physical effect, we need to consider explicitly the same realisation.

Interestingly, the best-fit values
also vary with the type and the level of PNG. One reason for this
could be the modified distribution of eigenvalues of the initial shear
field as found in the presence of primordial non-Gaussianity by
\cite{Lam}. 
The dependence of $D_B$ and $\beta$ on $f_{NL}$ was already found in
\cite{AC1} for the local type on non-Gaussianity. There, a linear
scaling in $f_{NL}$ modelled the best-fit values of $D_B$ and $\beta$
well. In the analysis presented here this is the case as well. Note, however, that due to
the small mass range covered by our simulations (minimum halo mass
is $5\times 10^{13}\,{\rm M_{\odot}}/h$ with 100 particles), the
uncertainties in the derived parameters are large and strongly correlated.

\begin{figure}[h]
\begin{center}
\includegraphics[scale=0.7]{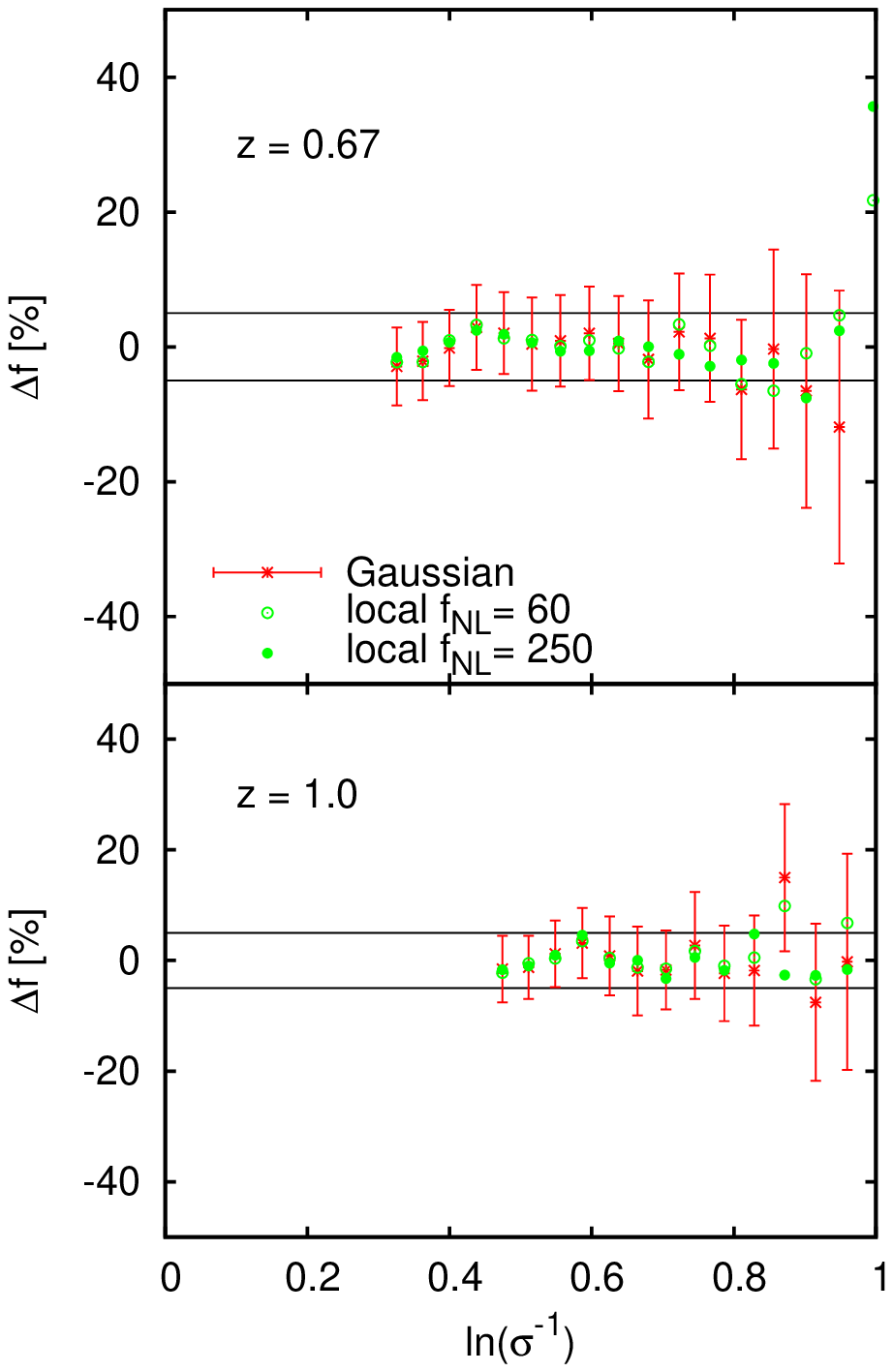}
\includegraphics[scale=0.7]{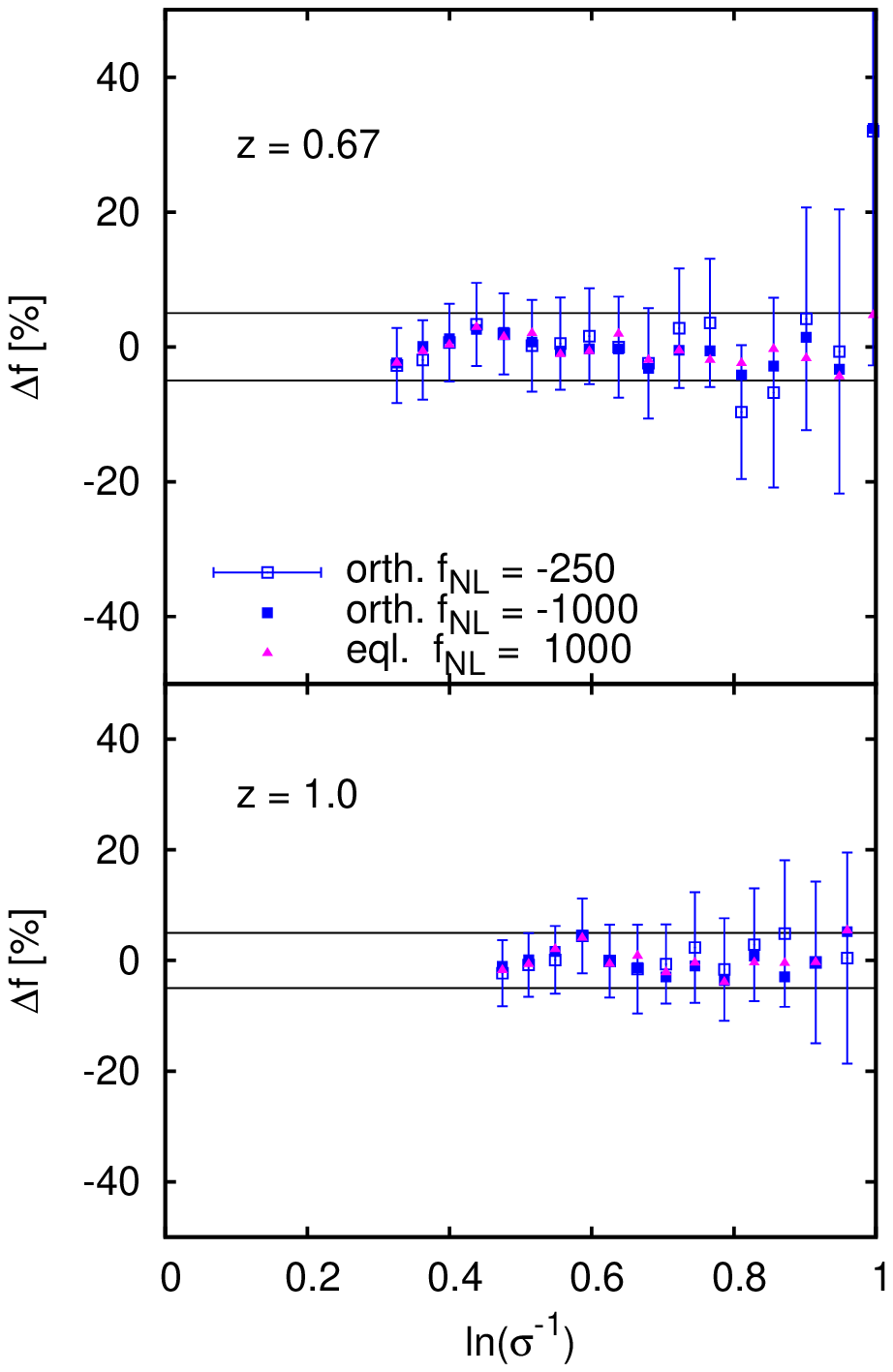}
\caption{Relative difference of the multiplicity functions $f(\sigma)$
  measured from N-body simulations and the corresponding best-fit DDB
  model predictions at $z=0.67,1$ for the same various types of
  primordial (non)-Gaussianity we show in
  Fig.~\ref{fig_ng_mf}. For clarity, we only show the error bars of one model in each panel. The error bars of the other models have virtually the same size.}\label{Fig8}
\end{center}
\end{figure}

The residuals of the best-fit DDB model and the multiplicity function of the simulation are shown in Fig.~\ref{Fig8} where we
can see once again that the barrier modelling is able to describe the
N-body simulation not only at $z=0$ but also at higher
redshifts\footnote{For the Gaussian case, see also \cite{CA2}}.

\section{Conclusion}

In summary, the DDB model is able to fit with high accuracy the halo
mass function from N-body simulations in various (non)-standard cosmological models with only two parameters. The non-standard models we consider here are a Ratra-Peebles quintessence dark energy and various amount and shape of PNG.
In the context of the DDB model, the two parameters, $D_B$ and
$\beta$, can be physically interpreted and linked to the PDF of the
initial critical overdensity around randomly selected particles
which will end up in halos. We also find a consistency in their evolution at different redshift showing that we are able for the first time to rewrite the problem of the non-universality of the mass function, as a problem of non-universality of the collapse barrier. A prediction of these parameters
from theory is still lacking just like an accurate prediction of the
PDF for critical overdensity at the center of mass of the proto-halo
is currently inconsistent with the ellipsoidal collapse model \cite{ST}
(e.g.\cite{Roberston,ARSC}). However, one can compute
those parameters from the the proto-halo distributions as we have shown in the case of the $\Lambda$CDM
simulations, for which we had good enough statistics to do so.
We found that, in general, the measured mean and variance of the critical density for collapse can be described reasonably well by a linear dependence on $S$.
For large masses, however, the scatter in the critical density for collapse measured from the initial conditions is larger than predicted by the best-fit linear model. It seems that even in the limit of $S\rightarrow 0$, the scatter does not go to zero as expected from the spherical collapse.
Nevertheless, the mass function predicted by the best-fit linear DDB model reproduces with high accuracy ($\sim 5\%$) the measured mass function from N-body simulations in the entire mass range.
Note, that the exact values of the barrier parameters depend on the specifics of the halo finder, the redshift, the cosmological model, and the shape and amplitude of the primordial non-Gaussianity. It might also depend on the environment of the halos as it is suggested by
\cite{Lee}. This could lead to further investigations: How does the
environment change the abundance of halos?  We hope to address this question in the near future.

\section*{Acknowledgement}
We are grateful to Ben Hoyle for proofreading this manuscript. This work is supported by the Trans-Regional Collaborative Research Center TRR 33 ``The Dark Universe'' of the Deutsche Forschungsgemeinschaft (DFG).

\end{document}